\documentstyle[aps,prl,preprint,floats,epsfig]{revtex}

\begin{document}

\preprint{\tighten\vbox{\hbox{\hfil CLNS 00/1691}
                        \hbox{\hfil CLEO 00-18}
}}

\title{Study of {\boldmath $\chi_{c1}$} and   {\boldmath $\chi_{c2}$} meson 
production in {\boldmath $B$} meson decays}

\author{CLEO Collaboration}
\date{\today}

\maketitle
\tighten

\begin{abstract} 
Using    a   sample   of
$9.7\times10^6$ $B  \overline B$ meson pairs collected with the CLEO
detector,    we study $B$ decays  to  the $\chi_{c1}$  and $\chi_{c2}$  charmonia  states, which 
are reconstructed via their radiative decays to $J/\psi$.
We first measure the  branching fraction for  inclusive 
$\chi_{c1}$ production  in  $B$  decays to be 
${\cal B}(B\to\chi_{c1}X)=(4.14\pm0.31\pm 0.40)\times10^{-3}$,
where the  first  uncertainty is statistical  and  the  second one  is
 systematic. 
 We  derive the branching fractions  for direct $\chi_{c1}$ and
$\chi_{c2}$  production in  $B$  decays by 
subtracting   the known contribution of  the decay chain 
$B\to\psi(2S)X$ with $\psi(2S)\to  \chi_{c1,2} \gamma$.  We obtain ${\cal
B}[B\to\chi_{c1}({\rm direct})X]=(3.83\pm0.31\pm 0.40)\times10^{-3}$.
  No  statistically   significant signal   for  $\chi_{c2}$
production is observed in either case. Using the Feldman-Cousins approach, 
we determine the 95\% confidence intervals to be $[0.2, 2.0]\times10^{-3}$ for ${\cal B}(B\to\chi_{c2}X)$,  $[0.0,1.7]\times10^{-3}$  for ${\cal B}[B\to\chi_{c2}({\rm direct})X]$, and $[0.00,0.44]$ for the ratio 
 $\Gamma[B \to \chi _{c2}({\rm direct})X]/\Gamma[B\to\chi_{c1}({\rm direc t})X]$.  
We also measure the branching ratio  $\Gamma[B \to \chi_{c2}({\rm direct})X_s]/\Gamma [B\to\chi_{c1}({\rm direc t})X_s]$ for different $X_s$ configurations by reconstructing  $B$ decays into  exclusive final states with $J/\psi$, $\gamma$, a kaon, and up to four pions. For all the  $X_s$ configurations 
we observe a strong $\chi_{c1}$ signal  yet no  statistically   significant 
$\chi_{c2}$ signal.
 We discuss how our results compare  with  theoretical predictions  in different models of charmonium production.
\end{abstract}
\newpage

{
\renewcommand{\thefootnote}{\fnsymbol{footnote}}

\begin{center}
S.~Chen,$^{1}$ J.~Fast,$^{1}$ J.~W.~Hinson,$^{1}$ J.~Lee,$^{1}$
D.~H.~Miller,$^{1}$ E.~I.~Shibata,$^{1}$ I.~P.~J.~Shipsey,$^{1}$
V.~Pavlunin,$^{1}$
D.~Cronin-Hennessy,$^{2}$ A.L.~Lyon,$^{2}$
% ...
\end{center}
 
\small
\begin{center}
$^{1}${Purdue University, West Lafayette, Indiana 47907}\\
$^{2}${University of Rochester, Rochester, New York 14627}\\
% ...
\end{center}

\setcounter{footnote}{0}
}
\newpage

The recent measurements  of charmonium production in
various high-energy physics reactions  have brought welcome  surprises
and challenged  our  understanding both of   heavy-quark production
and of  quarkonium  bound  state formation.
 The CDF and D0  measurements~\cite{CDF-D0-charmonia}  of a  
large  production rate for charmonium  at  high transverse momenta ($P_T$) 
were in sharp disagreement with the   then-standard color-singlet model.
The development of the  NRQCD factorization framework~\cite{Bodwin:1995jh} 
has put the calculations  of the inclusive charmonium 
production    on a rigorous  footing.
The high-$P_T$ charmonium
production rate at the Tevatron is now well understood in this formalism. 
The recent CDF 
measurement of  charmonium polarization~\cite{CDF-polarization}, 
however, appears to disagree with the NRQCD prediction.
The older color-evaporation model accommodates both the high-$P_T$
charmonium production rate and polarization measurements at the 
Tevatron~\cite{Amundson:1997qr}.
 
Inclusive $B$ decays to  charmonia offer another means by which theoretical predictions may be confronted with experimental data.
The color-singlet contribution, for example, is thought to be~\cite{Beneke:1999ks} a factor of 5--10
below   the observed inclusive $J/\psi$ production 
rate~\cite{Balest:1995jf}.
A  measurement of the 
$\chi_{c2}$-to-$\chi_{c1}$ production ratio in $B$ decays provides 
an  especially clean  test of  charmonium production models. 
       The        $V-A$         current        
 $\overline  c \gamma_{\mu}(1-\gamma_{5})c$ cannot create 
a  $c \overline c$ pair in a 
${^{2S+1}L}_J = {^3P}_2$ state,   therefore  the  decay
$B\to   \chi_{c2}  X$  is  forbidden     
 at   leading  order in $\alpha_s$  in the color-singlet 
model~\cite{chi-color-singlet}.
The importance of the 
color-octet mechanism for  $\chi_{c}$ production in $B$ decays  
was recognized~\cite{Bodwin:1992qr} even before the development of the 
NRQCD framework~\cite{Bodwin:1995jh}.
While the NRQCD calculations cannot yet produce sharp quantitative 
predictions for the $\chi_{c2}$-to-$\chi_{c1}$
production ratio in $B$ decays~\cite{Beneke:1999ks}, we can  
consider two limiting cases. If the   color-octet mechanism dominates 
in $B\to\chi_{cJ}X$ decays,
then the $\chi_{c2}$-to-$\chi_{c1}$ production ratio should be 5:3 because 
the color-octet contribution is proportional to $2J+1$.  In contrast, 
if the  color-singlet contribution dominates, then $\chi_{c2}$ production 
should be strongly suppressed relative to $\chi_{c1}$ production. 
The color-evaporation model  predicts the ratio to be 
$5:3$~\cite{Schuler:1999az}.

Our data  were collected at   the Cornell Electron Storage Ring
(CESR) with    two   configurations    of     the   CLEO detector
called CLEO~II~\cite{Kubota:1992ww}  and
CLEO~II.V~\cite{Hill:1998ea}.  The components of the CLEO detector
most relevant to this analysis are the charged particle tracking
system, the CsI electromagnetic calorimeter, the
time-of-flight system, and the muon chambers.
In CLEO~II the momenta of charged particles are measured in a tracking
system consisting of a 6-layer straw tube chamber,  a 10-layer
precision drift chamber, and a 51-layer main drift chamber, all
operating inside a 1.5 T  solenoidal magnet. The main drift chamber
also provides a measurement of the  specific ionization, $dE/dx$, used
for particle identification.  For  CLEO~II.V, the straw tube  chamber
was replaced  with a  3-layer silicon vertex detector, and the gas in
the main drift chamber was changed from an argon-ethane to a
helium-propane mixture. The muon chambers  consist of proportional
counters placed at increasing depths in  the steel absorber. 
 
 We use
9.2~$\rm fb^{-1}$ of $e^+e^-$ data taken at the $\Upsilon(4S)$
resonance    and 4.6~$\rm fb^{-1}$ taken  
60~MeV below the $\Upsilon(4S)$ resonance (off-$\Upsilon(4S)$ sample).  
Two thirds of the data  were collected with
the CLEO~II.V detector.  The simulated  event samples used in this
analysis were generated with a GEANT-based~\cite{GEANT} simulation of
the  CLEO detector response and  were processed in a manner similar to 
the  data.  

We reconstruct the $\chi_{c1,2}$  radiative decays to $J/\psi$. 
The branching fractions for 
the $\chi_{c1,2}\to J/\psi \, \gamma$ decays  are, respectively,   
$(27.3\pm1.6)\%$ and $(13.5\pm1.1)\%$, whereas the branching fraction  for the $\chi_{c0}\to J/\psi \, \gamma$ decay is only $(0.66\pm0.18)\%$~\cite{PDG}.
In addition, the $\chi_{c0}$ production rate in $B$ decays is expected 
to be smaller  than the $\chi_{c1,2}$ rates~\cite{Beneke:1999ks,Bodwin:1992qr}.
 We therefore do not attempt to measure  
$\chi_{c0}$ production  in this analysis. 

 The $J/\psi$ reconstruction procedure is described 
in Ref.~\cite{Avery:2000yh} and summarized here.  We reconstruct  both  $J/\psi \to \mu^+ \mu^-$ and  $J/\psi \to e^+ e^-$ decays, recovering the bremsstrahlung photons for the 
$J/\psi \to e^+ e^-$ mode.
We use the normalized invariant mass for the $J/\psi$ candidate selection (Fig.1 of Ref.~\cite{Avery:2000yh}).
For example, the
normalized  $J/\psi \to \mu^+ \mu^-$ mass is defined as  $[M(\mu^+
\mu^-)-M_{J/\psi}]/\sigma(M)$, where $M_{J/\psi}$ is the world average
value of the $J/\psi$ mass~\cite{PDG} and $\sigma(M)$ is the
expected mass resolution for that  particular  $\mu^+ \mu^-$
combination  calculated from track four-momentum     covariance matrices.
  We   require     the normalized
mass  to be between $-6$  and $+3$  for the $J/\psi  \to  e^+  e^-$ candidates and between  $-4$  and $+3$ for the  $J/\psi \to  \mu^+   \mu^-$ candidates.
The momentum of
the  $J/\psi$ candidates is required to be less than 2~GeV/$c$, which
is slightly above  the maximal $J/\psi$ momentum in  $B$ decays. 

 Photon candidates   for $\chi_{c1,2}\to J/\psi \, \gamma$  reconstruction 
must be detected in the central angular region  of the calorimeter ($|\cos\theta_{\gamma}|<0.71$), where our detector  has the best energy resolution.
Most of the  photons in  $\Upsilon(4S) \to B \overline  B$
events   come from $\pi^0$ decays. 
We therefore discard those photon candidates which, when paired with another
$\gamma$ in the event, produce  a  normalized $\pi^0 \to \gamma \gamma$  mass between
$-3$ and $+2$. 

\begin{figure}[htbp]
\centering                                            
\epsfxsize=170mm
\epsfbox{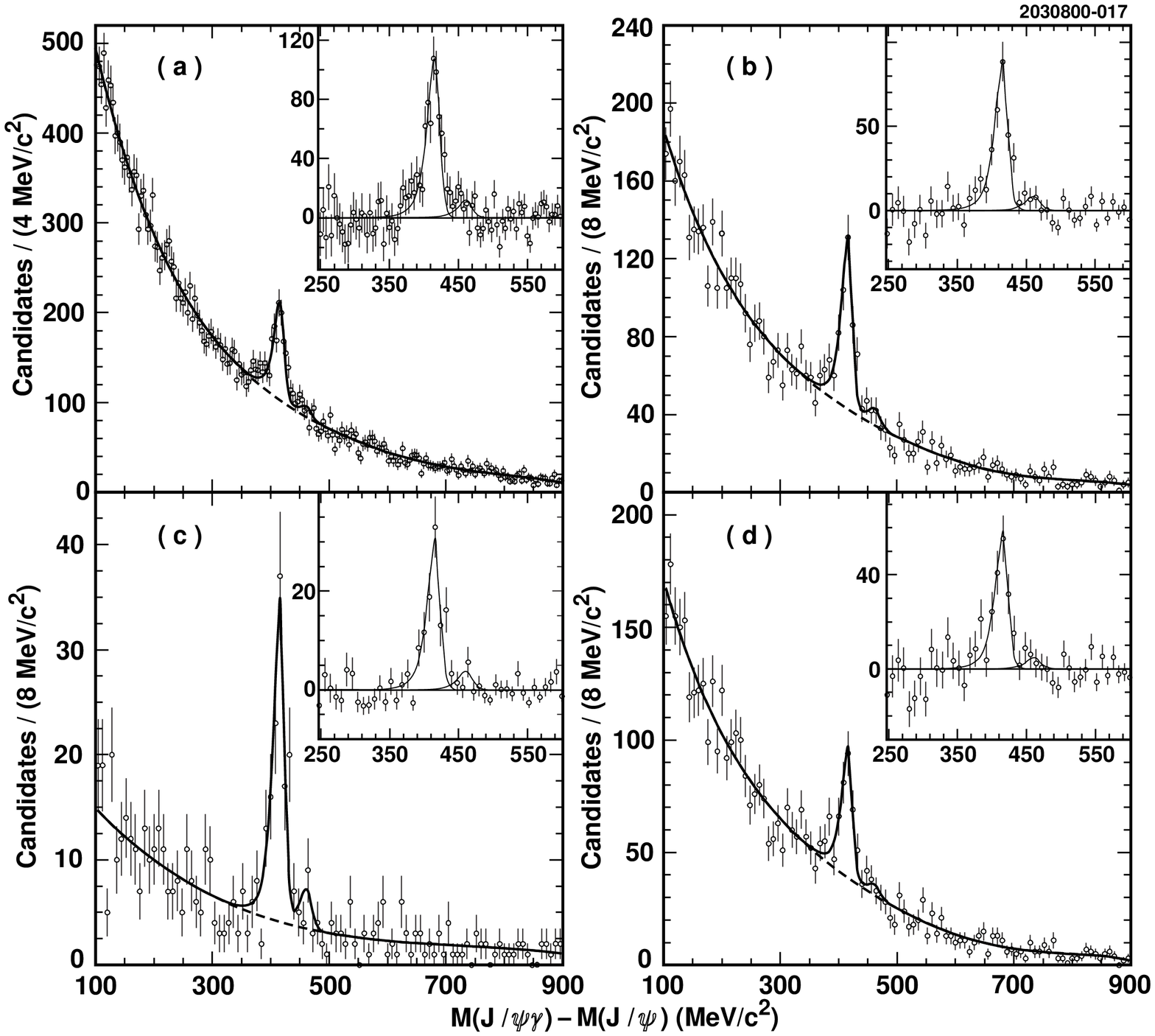}
\caption{ The  $M(J/\psi \gamma)-M(J/\psi)$ distribution in the $\Upsilon(4S)$ data (points with error bars). Plot (a) is for inclusive 
$J/\psi \gamma$ combinations, whereas plots (b), (c), and (d) are for 
those $J/\psi \gamma$ combinations that reconstruct 
to a $B \to J/\psi \gamma X_s$ decay with the $X_s$ composition corresponding to samples $A$, $B$, and $C$ described in the text. 
The fit function is shown by a solid line with the background component represented by a dashed line. The insets show the background-subtracted distributions 
with the $\chi_{c1}$ and $\chi_{c2}$ fit components  represented by a solid line.}
\label{fig:chi_incl_data_fit}
\end{figure}

In the first part of this work, called the inclusive analysis, we investigate 
$B\to \chi_{c1,2}X$ decays reconstructing only $J/\psi$ and $\gamma$. 
We determine the $\chi_{c1}$   and  $\chi_{c2}$  yields in 
a binned maximum-likelihood fit to the mass-difference distribution $M(J/\psi \gamma)-M(J/\psi)$ (Fig.~\ref{fig:chi_incl_data_fit}a), 
where $M(J/\psi)$ is the measured  mass of a $J/\psi$ candidate. 
The excellent electromagnetic calorimeter allows us to resolve the 
$\chi_{c1}$   and  $\chi_{c2}$ peaks. The $M(J/\psi \gamma)-M(J/\psi)$ mass-difference resolution is 8~MeV/$c^2$ and is dominated by the photon energy resolution. The bin width in the fit is  1~MeV/$c^2$. 
The background in the fit is  approximated by a 
5th-order Chebyshev polynomial, chosen as the minimal-order 
 polynomial  well fitting the background in a high-statistics sample of simulated $\Upsilon(4S) \to B \overline  B$ events.
  All the polynomial coefficients are allowed to float in the fit. 
The $\chi_{c1}$   and  $\chi_{c2}$ signal shapes are fit with 
 templates extracted from Monte Carlo simulation;  only  the template 
normalizations are free in the fit.
The    $\chi_{c1}$   and  $\chi_{c2}$   signal   yields in  the $\Upsilon(4S)$ data
  are  $N^{\rm ON}
(\chi_{c1})=672\pm47({\rm stat})$  and $N^{\rm ON}(\chi_{c2})=83\pm37({\rm stat})$.
The  $\chi_{c1}$   and  $\chi_{c2}$  yields in off-$\Upsilon(4S)$ data 
are both  consistent with zero:     
$N^{\rm  OFF}(\chi_{c1})=4\pm7({\rm stat})$   
and   $N^{\rm OFF}(\chi_{c2})=1\pm7({\rm stat})$. 
Subtracting the contributions from non-$B\overline B$ continuum 
events, we obtain the total inclusive $B\to \chi_{c1}X $   and  $B\to\chi_{c2}X$ event  yields 
$N(B\to \chi_{c1}X)=664\pm49({\rm stat})$ and $N(B\to \chi_{c2}X)=81\pm39({\rm stat})$.

Taking into account the systematic uncertainties associated with the fit, 
we determine  the $B\to \chi_{c2}X$ signal yield significance to be $2.0$ standard deviations ($\sigma$). Subtracting the known contribution of 
the decay chain 
 $B\to \psi(2S)X$ with $\psi(2S)\to\chi_{c2}\gamma$ and accounting for the 
associated systematic uncertainty,  
we likewise determine the  
 significance of the evidence for the decay  $B\to \chi_{c2}({\rm direct})X$ to be only  $1.4\sigma$. 

To calculate the branching fractions ${\cal B}(B\to \chi_{c1,2}X)$,
we use the measured 
signal yields $N(B\to \chi_{c1,2}X)$, the reconstruction efficiencies, the number of produced $B
\overline B$  pairs, and the daughter branching fractions.
The reconstruction efficiencies, determined from simulation, are  
$(25.7\pm0.2)\%$ for $\chi_{c1}$ and $(26.6\pm0.2)\%$ for $\chi_{c2}$, where 
the uncertainties are due to the size  
of our $B\to \chi_{c1,2}X$ simulation samples.
For the calculation of the rates  for the decays $B\to \chi_{c1,2}({\rm direct})X$, we make an assumption that the only other source of  $\chi_{c1,2}$  
production  in 
$B$ decays is the decay chain  $B\to \psi(2S)X$ with $\psi(2S)\to\chi_{c1,2}\gamma$. The 95\% confidence intervals are calculated using  the Feldman-Cousins approach~\cite{Feldman:1998qc}. 
The resulting branching fractions are listed in 
Table~\ref{tab:results}.
Taking into account correlations between   
the  uncertainties, we obtain the branching ratio 
$\Gamma [B\to\chi_{c2}({\rm direct})X]/\Gamma[B\to\chi_{c1}({\rm direct})X]=0.18\pm0.13\pm 0.04$; the 95\% CL upper limit  on the ratio is $0.44$.
\begin{table}[htbp]
\center
\caption{\small Branching fractions for inclusive 
$B$ decays to $\chi_{c1}$ and $\chi_{c2}$.}
\label{tab:results}
\begin{tabular}{llc} 
Branching fraction   &  Measured value    & 
95\% CL interval  \\ 
  & ($\times10^{-3}$)  &  ($\times10^{-3}$) \\ \hline
${\cal B}(B\to\chi_{c1}X)$ & $4.14\pm0.31\pm 0.40$ & --- \\ 
~~${\cal B}[B\to\chi_{c1}({\rm direct})X]$ & $3.83\pm0.31\pm0.40$ & --- \\ 
${\cal B}(B\to\chi_{c2}X)$ & $0.98\pm0.48\pm 0.15$ & $[0.2,2.0]$  \\
~~${\cal B}[B\to\chi_{c2}({\rm direct})X]$ & $0.71\pm0.48\pm 0.16$ & $[0.0,1.7]$ \\ 
\end{tabular}
\end{table}

The systematic uncertainties are listed in Table~\ref{tab:systematics}. 
The sources of the uncertainty  can be grouped into three categories:

{\it Fit procedure.---}
This category includes the uncertainties  due to our choice of 
 the signal and  background shapes as well as the   bin size.
To fit the $\chi_{c1}$   and  $\chi_{c2}$ signal, we use 
the templates extracted from simulation.  We therefore are sensitive  to 
imperfections in the  simulation of the photon  energy measurement.  
 The systematic uncertainties associated with the simulation of 
the calorimeter 
response are estimated by comparing the $\pi^0\to \gamma\gamma$ invariant mass lineshapes for inclusive 
$\pi^0$ candidates in the data and in  Monte Carlo  samples. Then  the 
$\chi_{c1}$   and  $\chi_{c2}$ templates are modified accordingly 
in order to determine the resulting uncertainty in the signal yields.
To estimate the uncertainty associated with the calorimeter  energy scale, 
we   shift  the  $\chi_{c1}$   and  $\chi_{c2}$ templates 
by $\pm0.6$~MeV/$c^2$ 
in the fit. The uncertainty due to  time-dependent  variations of the 
calorimeter  energy scale is small compared to the 
overall energy scale uncertainty.
To estimate the uncertainty due to the calorimeter  energy resolution,
we change the width of the $\chi_{c1}$   and  $\chi_{c2}$ templates by 
$\pm4\%$. The uncertainty in the background shape is probed by fitting the background with a  template extracted from  high-statistics samples of 
simulated $\Upsilon(4S) \to B \overline  B$ and non-$B \overline  B$
continuum events; only the template normalization, not its shape, is allowed to float in the fit. 

{\it Efficiency calculation.---}
This category includes the   uncertainties in   the number of produced $B
\overline B$  pairs,   tracking efficiency, photon   detection efficiency,  lepton  detection efficiency, and model-dependence and 
statistical uncertainty   of  the $B\to\chi_{c1,2}X$ simulation.
The $\chi_{c1,2}$ polarization 
affects the   photon energy spectrum. 
 We  define  the helicity angle
$\theta_h$  to  be  the  angle  between  the  $\gamma$  direction   in
$\chi_{c}$ frame and the  $\chi_{c}$  direction in  the $B$ frame.  We
assume a flat $\cos\theta_h$  distribution in our simulation.
 The systematic uncertainty associated with this assumption is estimated by comparing the reconstruction efficiencies in the Monte Carlo  samples with $I(\theta_h)\propto \sin^2\theta_h$  and 
$I(\theta_h)\propto \cos^2\theta_h$ angular distributions. Parity 
is conserved in the decays $\chi_{c1,2}\to J/\psi \gamma$, so   
the helicity angle distribution contains only  even powers of $\cos\theta_h$.
Another source of uncertainty is our modeling of the  $X$ system 
in the $B\to\chi_{c1,2}X$ simulation. Photon detection efficiency depends on the  assumed model  
through  the $\chi_c$ momentum spectrum and  the $\pi^0$ multiplicity of  the final state. In our  simulation, we assume that $X$ is either a single $K$ or 
one of the  higher $K$ resonances; we also include the decay chain  $B    \to  \psi(2S)  X$ with $\psi(2S)\to\chi_{c1,2}\gamma$. To estimate the systematic uncertainty, we compare the $\chi_{c}\to J/\psi \gamma$ detection efficiency extracted  using this sample with the efficiency in the sample where we assume that $X$ is either a  $K^{\pm}$ or $K^0_S\to\pi^+\pi^-$.

{\it  Assumed branching fractions.---}
This category includes the   uncertainties on  the external 
 branching fractions.   
 We use the following values of the daughter  branching fractions:
${\cal  B}(J/\psi \to \ell^+  \ell^-)=(5.894\pm0.086)\%$~\cite{Bai:1998di},
 ${\cal B}(\chi_{c1} \to     J/\psi \gamma)=(27.3\pm1.6)\%$~\cite{PDG}, 
and  ${\cal B}(\chi_{c2} \to J/\psi \gamma)=(13.5\pm1.1)\%$~\cite{PDG}.
In the calculation of ${\cal B}[B\to\chi_{c1,2}({\rm direct})X]$,  
we also  assume the following values: ${\cal B}(B\to \psi(2S) X)= (3.5\pm0.5)\times10^{-3}$~\cite{PDG},  ${\cal B}(\psi(2S) \to \chi_{c1} \gamma)=(8.7\pm0.8)\%$~\cite{PDG}, and ${\cal B}(\psi(2S)\to\chi_{c2} \gamma)=(7.8\pm0.8)\%$~\cite{PDG}. 
\begin{table}[htbp]
\center
\caption{\small Systematic uncertainties on  ${\cal B}(B\to\chi_{c1,2}X)$.}
\label{tab:systematics}
\begin{tabular}{lcc} 
Source of & \multicolumn{2}{c}{relative uncertainty in \%} \\ 
systematic  uncertainty & ${\cal B}(B\to\chi_{c1}X)$ & 
${\cal B}(B\to\chi_{c2}X)$  \\ \hline
Fit procedure   &  &  \\ 
~~$\gamma$ energy scale & $0.4$ & $5.6$ \\ 
~~$\gamma$ energy  resolution & $2.8$ & $6.9$  \\ 
~~Background shape & $1.8$ & $6.8$ \\
~~Bin size & $0.0$ & $1.9$ \\ 
Efficiency calculation   &  &  \\ 
~~$N(B \overline B)$ & $2.0$ & $2.0$ \\ 
~~Tracking efficiency & $2.0$ & $2.0$ \\ 
~~Lepton identification & $4.2$ & $4.2$ \\ 
~~Photon finding & $2.5$ & $2.5$ \\ 
~~Monte Carlo statistics & $0.7$ & $0.7$ \\ 
~~Model for $X$ in $B\to\chi_{c1,2}X$  & $3.3$ & $3.3$  \\
~~Polarization of $\chi_{c1,2}$ & $1.0$
& $1.0$ \\ 
Assumed branching fractions   &  &  \\ 
~~${\cal B}(\chi_{c1,2}\to J/\psi \gamma)$ & $5.9$ & $8.1$ \\ 
~~${\cal B}(J/\psi \to \ell^+ \ell^-)$ & $1.5$ & $1.5$ \\ 
~~${\cal B}(B\to\psi(2S)X)$\tablenotemark[1] & $1.1$ & $5.5$ \\ 
~~${\cal B}(\psi(2S)\to\chi_{c1,2}\gamma)$\tablenotemark[1]  & $0.7$ & $4.0$ \\ 
\end{tabular}
\flushleft \tablenotetext[1]{Contributes only to uncertainty on ${\cal B}[B\to\chi_{c1,2}({\rm direct})X]$.}
\end{table}

In the second part of this work, called the $B$-reconstruction analysis,
we employ the $B$-reconstruction technique similar to the one  
developed for the $b \to s \gamma$ rate measurement~\cite{b-to-s-gamma}.
We still extract  $\chi_{c1}$ and $\chi_{c2}$ signal yields from a fit 
to  $M(J/\psi \gamma)-M(J/\psi)$ distribution, but  we select 
only  those $J/\psi \gamma$ combinations  that reconstruct 
to a $B \to J/\psi \gamma X_s$ decay. This  $B$-reconstruction technique is 
used to suppress backgrounds and allows us to probe the composition of the $X_s$ system 
accompanying $\chi_{c1,2}$ mesons.
We extract the branching ratio ${\cal R}(\chi_{c2}/\chi_{c1}) \equiv \Gamma[B\to\chi_{c2}({\rm direct})X_s]/\Gamma [B\to\chi_{c1}({\rm direct})X_s]$ for the following three $X_s$ configurations:
\begin{enumerate}
\item {\it Sample $A$}.--- $X_s$ is reconstructed as a kaon ($K^+$ or $K^0_S\to\pi^+\pi^-$) with 0 to 4 pions, one of which can be a $\pi^0$. We consider 21 possible $X_s$ modes as well as the charge conjugates of these modes.  
\item {\it Sample $B$}.--- $X_s$ is reconstructed as a single kaon or $K^*(892)$. 
A $K\pi$ combination is a $K^*$ candidate  if $|M(K\pi)-M_{K^*}|<75$~~MeV/$c^2$, where $M_{K^*}$ is the world average $K^*(892)$ mass~\cite{PDG}.
\item {\it Sample $C$}.--- $X_s$ is reconstructed as 
a kaon with 1 to 4 pions, but not as a 
$K^*(892)$ candidate ($|M(K\pi)-M_{K^*}|>200$~~MeV/$c^2$).
\end{enumerate}
Thus samples $B$ and $C$ are subsets of $A$. To an excellent approximation, 
sample  $A$ is a sum of $B$ and $C$.
With  sample $A$, we try to reconstruct as many $B\to J/\psi \gamma X_s$
decays as possible. Dividing  sample $A$ into subsamples $B$ and $C$, 
we also probe the dynamics of the  $B\to\chi_{c1,2}X_s$ decays.
If the dominant production mechanisms for $\chi_{c1}$ and $\chi_{c2}$ are different, color-singlet mechanism for $\chi_{c1}$ and   color-octet for $\chi_{c2}$, then it is natural to expect that 
 $\chi_{c2}$, in comparison with $\chi_{c1}$, is more often  accompanied by 
 multi-body $X_s$ states rather than a single $K$ or $K^*$.
Thus the measured $\chi_{c2}$-to-$\chi_{c1}$ production ratio might be quite different for samples $B$ and $C$.

We require  that  the charged kaon and pion  
candidates  have, if available, 
$dE/dx$ and time-of-flight  measurements that   lie within  $3\sigma$ 
of the expected values. The $dE/dx$ measurement is required for 
kaons, but used only if available for pions. The  time-of-flight measurement 
is used only if available.
The  $K^0_S \to \pi^+\pi^-$ candidates are selected   from  pairs of
tracks  forming   displaced vertices.
We require the absolute value of the  normalized  $K^0_S \to \pi^+\pi^-$  mass 
to be less than 4 and  perform a fit constraining  the mass of each $K^0_S$
candidate to the world average value~\cite{PDG}.  
Photon candidates   for $\pi^0
\to \gamma \gamma$ decays are required to have an energy of at least
30~MeV in the central region and 
at least 50~MeV in the endcap  region ($0.71<|\cos\theta_{\gamma}|<0.95$) 
of the calorimeter.  
We   require the  absolute value of the normalized  $\pi^0
\to \gamma \gamma$ mass to be less than 3 and  
perform a fit constraining  the mass
of each $\pi^0$   candidate to the world average value~\cite{PDG}.  
The  $J/\psi$ four-momentum used in 
$B\to J/\psi \gamma X_s$ reconstruction is obtained by performing  
a fit constraining   the $J/\psi$ candidate mass  to  the world average value~\cite{PDG}.

 The $B$  candidates are  selected by means of two  observables. The
first  observable  is the difference between the energy of  the $B$
candidate and the  beam  energy,  $\Delta E  \equiv E(B) - E_{\rm
beam}$. The average $\Delta E$ resolution varies from 12 to 17~MeV 
depending on the $B$-reconstruction mode. The   second
observable   is  the     beam-constrained  $B$    mass,
$M(B)\equiv\sqrt{E^2_{\rm beam}-p^2(B)}$, where $p(B)$ is 
the $B$ candidate momentum.  The average $M(B)$ resolution 
is 2.7~MeV/$c^2$ and is  dominated by the beam energy spread.
We use the normalized  $M(B)$ and  $\Delta E$  
variables and require $|\Delta E|/\sigma(\Delta E)<3$ and  
$|M(B)-M_{B}|/\sigma(M)<3$, where $M_{B}$ is the nominal $B$ meson
mass. 
The fit to  $M(J/\psi \gamma)-M(J/\psi)$ distribution is then performed in the 
same manner as in the inclusive analysis. We still use a 
5th order Chebyshev polynomial to fit the background for 
samples $A$ and $C$, but we reduce the order of the 
polynomial to 3 for the low-statistics sample $B$. The fits are shown in
Fig.~\ref{fig:chi_incl_data_fit} and   
the  $\chi_{c1}$ and  $\chi_{c2}$ signal yields are listed in Table~\ref{tab:b-reconstruction-results}. The $B$-reconstruction technique renders negligible 
the contribution from non-$B\overline B$ continuum  events.
 We finally subtract the $\psi(2S)\to\chi_{c1,2}\gamma$ feeddown to obtain the 
rates for direct $\chi_{c1,2}$ production in $B$ decays. 
 For all three $X_s$ configurations, we observe a strong $\chi_{c1}$ signal  yet no  statistically   significant 
signal   for  direct $\chi_{c2}$ production (Table~\ref{tab:b-reconstruction-results}).  To  calculate the branching ratio ${\cal R}(\chi_{c2}/\chi_{c1})$, 
we multiply  the ratio of the feeddown-corrected  $\chi_{c1,2}$   yields by  
the  reconstruction efficiency ratio ${\cal E}(\chi_{c1})/{\cal E}(\chi_{c2})$ 
and by 
the branching ratio $\Gamma (\chi_{c1}\to J/\psi\gamma)/\Gamma (\chi_{c2}\to J/\psi\gamma)$.
The efficiency of the $B$-reconstruction  depends on the composition of the   $X_s$ system. We assume that the $X_s$ system composition is the same for  $\chi_{c1}$ and $\chi_{c2}$ production. 
From our simulation we determine ${\cal E}(\chi_{c1})/{\cal E}(\chi_{c2})\simeq0.93$ for all three $X_s$ configurations.  
The resulting  $\chi_{c2}$-to-$\chi_{c1}$ production ratios are listed in 
 Table~\ref{tab:b-reconstruction-results}.
\begin{table}[htbp]
\center
\caption{\small Results for each of the three $X_s$ configurations 
used in $B\to J/\psi \gamma X_s$ reconstruction.  
The $\chi_{c1}$ and $\chi_{c2}$ event yields with associated statistical uncertainties are listed in lines 1 and 2. Line 3 contains  the significance of the $B\to\chi_{c2}({\rm direct})X_s$ signal  with statistical and systematic uncertainties taken into account. Lines 4 and 5 contain the measured value and 95\% confidence interval for the  branching ratio ${\cal R}(\chi_{c2}/\chi_{c1}) \equiv \Gamma [B\to\chi_{c2}({\rm direct})X_s]/\Gamma[B\to\chi_{c1}({\rm direct})X_s]$, determined with an assumption that the $X_s$  system composition is the same for  $\chi_{c1}$ and $\chi_{c2}$ production.}
\label{tab:b-reconstruction-results}
\begin{tabular}{lccc} 
               & Sample $A$ & Sample $B$  & Sample $C$ \\ \hline
 $N(B\to\chi_{c1}X_s)$ & $279\pm25$ & $96\pm12$ & $183\pm22$ \\  
 $N(B\to\chi_{c2}X_s)$ & $31^{+18}_{-17}$ & $13.9^{+7.0}_{-6.2}$ & $18\pm16$ \\ Significance of  $B\to\chi_{c2}({\rm direct})X_s$ & $1.2\sigma$ & $2.0\sigma$ & $0.6\sigma$ \\ 
${\cal R}(\chi_{c2}/\chi_{c1})$ & $0.18\pm0.12\pm0.09$ & $0.27^{+0.15}_{-0.13}\pm0.05$ & $0.14\pm0.18\pm0.14$ \\
95\% CL interval for ${\cal R}(\chi_{c2}/\chi_{c1})$ & $[0.00,0.48]$ &  $[0.04,0.58]$ &  $[0.00,0.59]$ \\
\end{tabular}
\end{table}

The systematic uncertainties for the $B$-reconstruction analysis are listed in Table~\ref{tab:b-reconstruction-syst}.
The sources of  uncertainty  can be grouped into the following  four 
categories:

{\it Fit procedure.---} As in the inclusive analysis, we estimate the 
uncertainties in the signal and background shapes. 
We  shift  the  $\chi_{c1,2}$   templates 
by $\pm0.6$~MeV/$c^2$ and vary their widths by $\pm4\%$.
The  requirement on  $\Delta E$ in $B \to J/\psi \gamma X_s$ 
reconstruction truncates the low-side tail of  the  $\chi_{c1,2}$ shapes. We estimate the uncertainty due to this effect
by using the $\chi_{c1,2}$   templates  obtained from the simulation with a requirement that the measured  $\chi_{c}$ energy is within $3\sigma$ of the 
generated value.
The uncertainty in the background shape dominates the 
fit procedure uncertainty.
To probe this uncertainty, we fit the  background with different  
templates, allowing only the template normalization, not its shape, to float in the fit. 
 One  template is 
extracted from simulation separately for each of the samples $A$, $B$, and $C$.
Another template, the same  for all three $X_s$ configurations, is 
the background shape from the inclusive analysis
(Fig.~\ref{fig:chi_incl_data_fit}a).

{\it $\psi(2S)$ subtraction.---}  
The sources of the systematic uncertainty associated with the  $\psi(2S)$-feeddown subtraction  include ${\cal B}(B\to \psi(2S) X)$, ${\cal B}(\psi(2S) \to \chi_{c1,2} \gamma)$, the 
size of our $B\to \psi(2S) X$  simulation sample, 
and  the composition of $X$ in $B\to \psi(2S) X$ decays. 
To estimate the uncertainty due to our  model of the $X$ system composition in 
the $B\to \psi(2S) X$ simulation, we check whether 
the data and the simulation agree on the ratio of 
$\psi(2S) \to \ell^+ \ell^-$ event yields 
obtained in the inclusive reconstruction  and after 
the $B\to \psi(2S) X_s$ reconstruction.   
This category  also  includes the uncertainties that would  have canceled  for the  ratio ${\cal R}(\chi_{c2}/\chi_{c1})$ were it not for the $\psi(2S)$-feeddown subtraction. 
These sources of uncertainty are 
${\cal B}(J/\psi \to \ell^+ \ell^-)$, 
$N(B \overline B)$, tracking, photon finding, and 
lepton identification. 

{\it ${\cal E}(\chi_{c2})/{\cal E}(\chi_{c1})$.---}
We assume that the $X_s$ system in $B\to\chi_{c1,2}X_s$ is the same for 
$\chi_{c1}$ and $\chi_{c2}$. We do not assign any uncertainty for this assumption. The remaining sources of uncertainty are  the  $\chi_{c1,2}$ polarization and  the statistics of the $B\to\chi_{c1,2}X_s$ simulation  samples. 

{\it ${\cal B}(\chi_{c1,2} \to J/\psi \gamma)$.---} Our measurement 
depends on the ratio 
$\Gamma(\chi_{c1} \to J/\psi \gamma)/\Gamma (\chi_{c2} \to J/\psi \gamma)$ and its uncertainty.
\begin{table}[htbp]
\caption{ The absolute systematic uncertainties on the branching ratio ${\cal R}(\chi_{c2}/\chi_{c1})$ for each of the three $X_s$ configurations 
used in  $B\to J/\psi \gamma X_s$ reconstruction.}
\begin{center}
\begin{tabular}{llll} 
\label{tab:b-reconstruction-syst}
  & \multicolumn{3}{c}{uncertainty on ${\cal R}(\chi_{c2}/\chi_{c1})$}   \\  
Source of uncertainty   & Sample $A$  & Sample $B$ & Sample $C$\\ \hline 
Fit procedure   & $ 0.084$  & $ 0.039$ & $ 0.142$ \\ 
$\psi(2S)$ subtraction 
                & $0.007$  & $ 0.001$ & $0.006$ \\ 
${\cal E}(\chi_{c1})/{\cal E}(\chi_{c2})$
                & $ 0.003$ & $ 0.006$ & $0.003$ \\
${\cal B}(\chi_{c1,2} \to J/\psi \gamma)$
                & $ 0.022$ & $ 0.026$ & $0.019$ \\  \hline 
Added in quadrature   & $0.09$  & $ 0.05$ & $0.14$  \\ 
\end{tabular}
\end{center}
\end{table}

In conclusion, we have measured the branching fractions for  inclusive 
$B$ decays   to  the  $\chi_{c1}$  and $\chi_{c2}$  charmonia  states.
Our measurements are consistent with and supersede  the previous 
CLEO results~\cite{Balest:1995jf}. We have also studied $B\to\chi_{c1,2}X_s$
decays, reconstructing $X_s$ as a kaon and up to four pions.
In this way, we have measured the branching ratio $\Gamma[B\to\chi_{c2}({\rm direct})X_s]/\Gamma[B\to\chi_{c1}({\rm direct})X_s]$ for three
$X_s$ configurations. 
In all the cases, we observe strong $\chi_{c1}$ signal 
yet no  statistically   significant signal    
for  $\chi_{c2}$ production.
Our measurement of the $\chi_{c2}$-to-$\chi_{c1}$ production 
ratio in $B$ decays is consistent with the prediction of the color-singlet model~\cite{chi-color-singlet} and disagrees with the color-evaporation model~\cite{Schuler:1999az}.
In the NRQCD framework, our measurement suggests that the color-octet 
mechanism does not dominate in $B\to\chi_{c}X$ decays.
           
We gratefully acknowledge the effort of the CESR staff in providing us with
excellent luminosity and running conditions.
This work was supported by 
the National Science Foundation,
the U.S. Department of Energy,
the Research Corporation,
the Natural Sciences and Engineering Research Council of Canada, 
the A.P. Sloan Foundation, 
the Swiss National Science Foundation, 
the Texas Advanced Research Program,
and the Alexander von Humboldt Stiftung.

\end{document}